\documentclass[preprint,review,14pt]{elsarticle}
\usepackage{lineno,hyperref}
\usepackage{graphicx}
\usepackage{epsfig}
\usepackage{amsmath,amsfonts, amsthm, mathrsfs, amssymb, bbm}
\usepackage{times}
\usepackage{subfigure}

\newcommand{\be}{\begin{equation}}
\newcommand{\ee}{\end{equation}}
\newcommand{\bea}{\begin{eqnarray}}
\newcommand{\eea}{\end{eqnarray}}

\newcommand{\nd}{\noindent}

\journal{arXiv preprint}

\begin{document}

\begin{frontmatter}

\title{Immune checkpoint therapy modeling of PD-1/PD-L1 blockades reveals subtle difference in their response dynamics and potential synergy in combination}

\author[DartMath]{Kamran Kaveh}
\ead{kkavehma@gmail.com}
\author[DartMath,DartBMDS]{Feng Fu\corref{sdp}}
\ead{fufeng@gmail.com}

\address[DartMath]{Department of Mathematics, Dartmouth College, Hanover, NH 03755, USA}

\address[DartBMDS]{Department of Biomedical Data Science, Geisel School of Medicine at Dartmouth, Lebanon, NH 03756, USA}

\cortext[sdp]{Corresponding author at: 27 N. Main Street, 6188 Kemeny Hall, Department of Mathematics, Dartmouth College, Hanover, NH 03755, USA. Tel: +1 (603) 646 2293, Fax: +1 (603) 646 1312}

\begin{abstract}
Immune checkpoint therapy is one of the most promising immunotherapeutic methods that are likely able to give rise to durable treatment response for various cancer types. Despite much progress in the past decade, there are still critical open questions with particular regards to quantifying and predicting the efficacy of treatment and potential optimal regimens for combining different immune-checkpoint blockades. To shed light on this issue, here we develop clinically-relevant, dynamical systems models of cancer immunotherapy with a focus on the immune checkpoint PD-1/PD-L1 blockades. Our model allows the acquisition of adaptive immune resistance in the absence of treatment, whereas immune checkpoint blockades can reverse such resistance and boost anti-tumor activities of effector cells. Our numerical analysis predicts that anti-PD-1 agents are commonly less effective than anti-PD-L1 agents for a wide range of model parameters. We also observe that combination treatment of anti-PD-1 and anti-PD-L1 blockades leads to a desirable synergistic effect. Our modeling framework lays the ground for future data-driven analysis on combination therapeutics of immune-checkpoint treatment regimes and thorough investigation of optimized treatment on a patient-by-patient basis.
\end{abstract}


\end{frontmatter}


\section{Introduction}
Immune system is shown to have the potential to activate a response that can eradicate a tumor \cite{ribas2015releasing,ribas2018cancer,sharma2015immune}. This has been identified long before a detailed understanding of the mechanisms and components of immune response were understood. Physicians had noticed that solid tumors regressed or even disappeared in patients with skin infections such as Erysipelas caused by Streptococci \cite{kucerova2016spontaneous}. While immune system has a vast potential to mount a response against a malignant tumor, in most cases tumor evolves and gains the capability to escape the anti-tumor response. Cancer cells gain the capability to become `invisible' (not recognized by) to the immune cells. Alternatively, they might be able to suppress or reverse the immune response \cite{ribas2015adaptive, kalbasi2019tumour}. This can be due to mutations or epigenetic/adaptive changes in cancer cells that increase the expression of immune suppressive pathways.

Immunotherapeutic techniques in cancer are, in essence, a category of methods to reverse the above mechanism of immune resistance (or escape) established by tumor cells. If, through therapeutic interventions, one can stop tumor cells from evading immune response, or inhibit tumor cells from suppressing it, then the immune system, in principle, is capable of eradicating the tumor population. Arguably, the most promising strategies in immunotherapy has been the development of immune checkpoint blockade antibodies \cite{ribas2018cancer, sharma2015immune, wei2018fundamental,littman2015releasing}.  Immune checkpoint blockade therapies are a new therapeutic method to make tumor cells `visible' to the immune cells and re-activate and strengthen tumor-specific immune response. This is achieved by blocking the signals/proteins or corresponding ligands that contribute in the immune recognition/activation or immune suppression pathways.

There have been several classes of immunotherapy drugs that function as immune checkpoint blockades. Antibodies that block cytotoxic T lymphocyte-associates protein 4 (CTLA-4) or programmed cell death-1 (PD-1) or its ligand-1 (PD-L1) are the most well-known immune checkpoint treatments. There has been success in patient survival in different cancers mainly, melanoma, non-small cell lung cancer and non-hodgekins lymphoma, among others. Currently, there are several immune checkpoint blockades that are approved by the FDA , such as Ipilimumab, Tremelimumab (anti-CTAL-4), Nivolumab (anti-PD-1) and Avelumab (anti-PD-L1). There are more immune checkpoint antibodies under development for a broader range of tumor types \cite{wei2018fundamental, tang2018comprehensive}. 

The field of immune checkpoint therapy has joined the ranks of surgery, radiation, chemotherapy, and targeted therapy as a pillar of cancer therapy. These drugs represent a radical and disruptive change in cancer therapy. With the exception of anti-PD-L1 agents, these drugs do not directly attack tumor cells. But instead, they target immune cells and immune system at large. More importantly, instead of activating the immune system, they suppress inhibitory pathways that induce immune resistance and block effective anti-tumor response by the immune system. Immune checkpoint therapy, with anti-CTLA-4 having longer follow-up than other agents, leads to durable clinical responses that can last a decade and more, but only in a fraction of patients \cite{sharma2015future}.

Unlike anti-PD-1 and anti-CTLA4 antibodies, anti-PD-L1 agents do target tumor cells. While anti-PD-1 antibodies block the pathways that suppress anti-tumor response in T-cells, anti-PD-L1 antibodies block similar pathways in tumor cells. They make tumor cells vulnerable to anti-tumor response. Successes have been seen in preclinical and clinical trials based on the combination of anti-CTLA4 and anti-PD-1 agents~\cite{wei2017distinct,wei2019combination}. As such, there are high expectations that a highly effective immunotherapeutic strategy can be devised using potential combinations of these different agents. There are ongoing studies to identify immune biomarkers with which one can predict the efficacy of treatment for select patients~\cite{willis2015immune}. But the complexity fo the immune system has made this task more difficult so far.

For an adaptive anti-tumor response to initiate, cytotoxic T cells that are responsible for killing tumor cells need to be activated in the first place. T cell activation is commonly specific to tumors with particular genetic makeup. Cytotoxic T cells are denoted with their protein marker CD8 (vs CD4 for helper T-cells). Upon interaction with tumor cells -- in the tumor microenvironment -- T cells can be activated. This also can happen indirectly when T cells encounter other immune cells that carry the tumor antigen. These are so-called antigen presenting cells or APC. APC's are commonly dendritic cells that have absorbed antigens from dying tumor cells, but they can be other immune cell type as well. \cite{wei2018fundamental,chaplin2010overview}.  

\begin{figure}
\centering
  \includegraphics[width=0.6\textwidth]{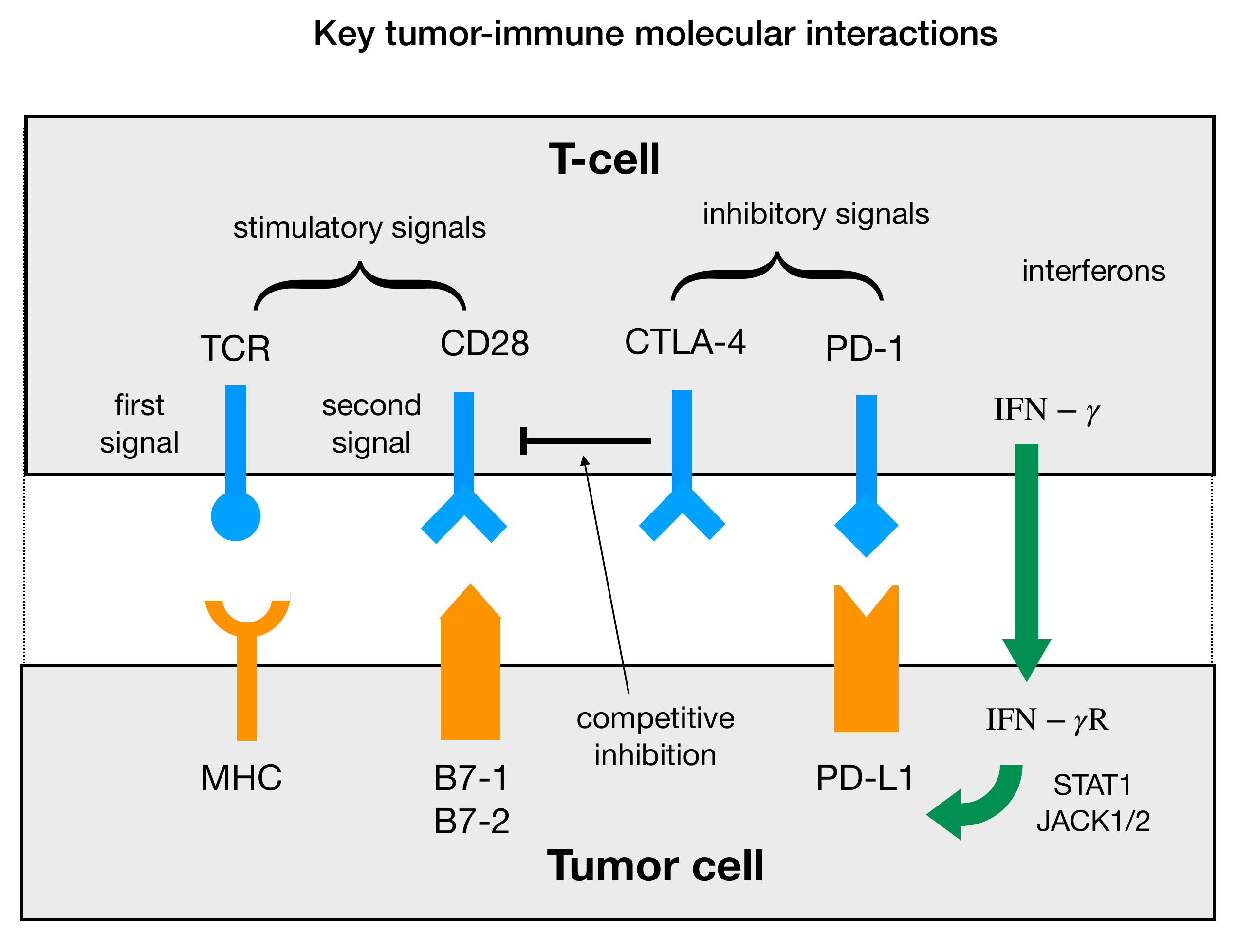}
\caption{Schematic illustration of key tumor-immune molecular interactions. TCR/MHC is the first immune activation signal and CD28/B7-1/2 is the second signal. CTLA-4 competitive inhibition with CD28 and PD-1/PD-L1 engagement are inhibitory signals for T cells.}
\label{immune0}
\end{figure}

Immune activation is done through a two-signal model (see Fig.~\ref{immune0}). The first and main signal is the binding of T cell receptor (TCR) by the MHC protein on APC or tumor cells. Second signals, or co-stimulating signals, are commonly CD28 on T cells that interact with B7-1/2 on tumor cells. One major function of a second signal is to stabilize the first signal binding. Upon activation T cells undergo clonal expansion with help of the IL-2 cytokines, which is a growth and differentiation factor. Beside clonal expansion of T cells, tumor-specific  APC cells are amplified in numbers as well. This is done through an intricate set of mechanisms that involve CD8 and CD4 T cells as well regulatory T cell (Tregs). This response can last by establishing memory T cells \cite{nicholson2016immune}.  
 
The mechanism of adaptive (or constitutive) immune resistance by tumor is understood by the over-expression of PD-L1 ligand on tumor cells. Upon engaging with its PD-L1 ligand, PD-1 acts as an immune suppressive signal \cite{gibbons2017functional}. When tumor antigen-specific T cells recognize their cognate antigen expressed by cancer cells, signaling through TCR leads to production of interferons and expression of regulatory receptors such as PD-1. Interferons tune the expression of immune-suppressive factors on tumor cells, including PD-L1 ligand. Upon engaging with PD-L1, PD-1 protein on a T cell works towards suppression of the adaptive immune response \cite{wu2019pd}. Cancer cells use these adaptive immune suppression programs that are set to limit the immune and inflammatory responses, to their benefit. The mechanism of immune resistance can be constitutive due to activation of oncogenic pathways that also lead to expression of PD-L1 ligand \cite{parsa2007loss, akbay2013activation, atefi2014effects}. However, it seems that adaptive interferon-inducible expression of PD-L1 is more common than the constitute expression in many cancers \cite{taube2012colocalization,tumeh2014pd}. These key tumor-immune molecular interactions are depicted in Fig.~\ref{immune0}. 

Anti-PD-1/PD-L1 immune checkpoint blockade therapies are aimed to reverse the immune resistance mechanism by making tumor cells visible to immune cells. This effectively leads to re-launching an immune response that is suppressed by PD-L1+ tumor cells. As described above, interaction of programmed cell death-1 (PD-1) protein with its ligand on a tumor cell sets off inhibition programs that suppresses the adaptive immune response. Immune checkpoint PD-1/PD-L1 blockades effectively inhibit PD-1/PD-L1 engagement. 

As aforementioned, another category of immune checkpoint blockades targets the CTLA-4 protein. The CTLA-4 is homologous to the T cell co-stimulatory protein, CD28, and both molecules bind B7-1/2 proteins. CTLA4 can outcompete CD28 in binding. It transmits an inhibitor signal for T cell activation, whereas CD28 transmits a stimulatory signal. CTLA-4 inhibition happens in T cell priming sites, i.e., lymph nodes~\cite{ribas2018cancer}. Blocking CTLA-4 on T cells increases the immune response and amplifies the activation process. CTLA-4 blockade, furthermore, increases T cell motility and renders more T cells to move into the tumor microenvironment as well. Such an increase in motility can significantly help in T cell infiltration process and anti-tumor response~\cite{pentcheva2014cytotoxic}. 

In what follows, we aim to address fundamental open questions regarding the dynamics of the immune checkpoint blockades using a quantitative modeling approach. Specifically, potential biomarker mechanisms behind the choice of therapeutic strategies 
using either of anti-PD-1/PD-L1 agents or both (monotherapy versus combination therapy) are not well-understood~\cite{linhares2019therapeutic,chen2018sequential,shen2018efficacy}. Moreover, there are still critical open questions with particular regards to quantifying and predicting the efficacy of treatment and potential optimal regimens for combining different immune checkpoint inhibitors. Our approach follows the mathematical oncology paradigm~\cite{basanta2012investigating,basanta2012investigating,szeto2019integrative,altrock2015mathematics,brady2019mathematical}, and our dynamical systems modeling can be used to inform the rational and personalized development of cancer immunotherapy using checkpoint blockades and their potential combinations in order to improve response rate and reduce resistance.

\section{Dynamical systems model of adaptive immune resistance through PD-1/PD-L1 axis}

A minimal dynamical model of immune-tumor interaction is recently suggested in Ref.~\cite{garcia2020cancer}. The model considers a population of tumor cells, $y$ and immune cells (effector cells $x$), in the tumor microenvironment. We assume a constant supply of (primed) effector cells from lymph nodes ($\lambda$). Upon interaction with tumor cells (either direct or indirect through APC cells) immune cells receive stimulatory signals to increase their proliferation. We assume this to happen with a rate $k$ per unit time and per capita. Tumor cells are killed with a rate $m$ upon interaction with effector cells. Assuming a logistic growth for tumor population in the absence of immune interaction, we have~\cite{garcia2020cancer}:

\begin{align}
\dot{x} &= \lambda - \mu \cdot x + k \cdot x y \nonumber\\
\dot{y} &= a y \cdot \big( 1 - b y \big) - m \cdot x y
\end{align} 
  
\nd where $a$ is the linear growth rate (fitness) of tumor cells, and $b$ is the inverse carrying capacity. $\mu$ is the average death rate of effector cells. The solution for the above equation shows bi-stability as the treatment parameters changes, and one can show what ranges of $m$ and $k$ values might represent a successful treatment. 

In the following we generalize the above model for immune checkpoint combination therapies. In this paper, we focus on anti-PD-1 and anti-PD-L1 antibodies and their effects on suppressing the immune resistance. 
  
\begin{figure}
\centering
  \includegraphics[width=0.6\textwidth]{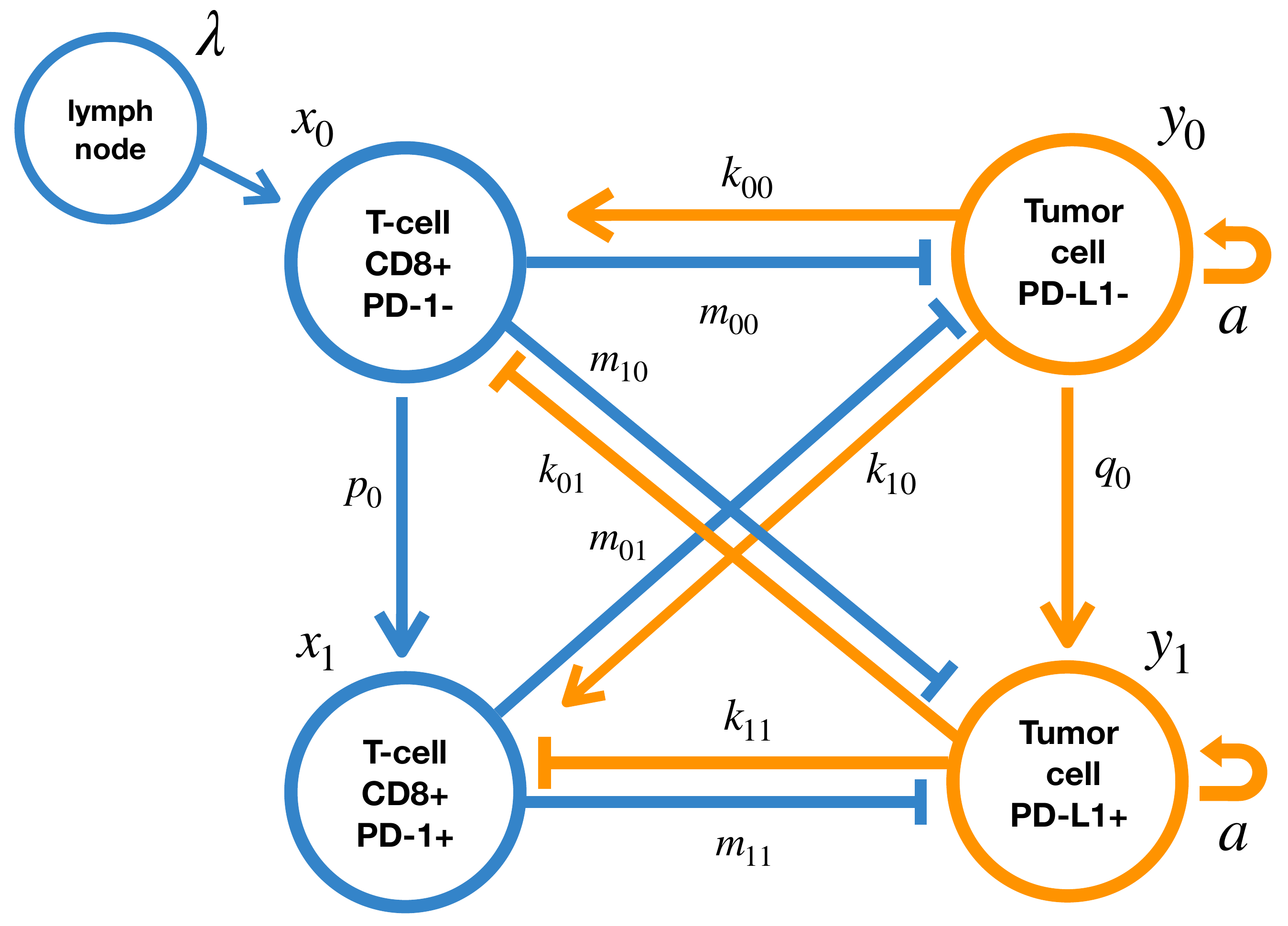}
\caption{
Model scheme of tumor-immune interactions in the microenvironment. Population of effector cells are divided into two subpopulation of PD-1+ and PD-1- types. Two tumor subpopulations of PD-L1+ and PD-L1- are distinguished in a similar fashion. Tumor and immune cell subpopulations interact based on matrices $k_{ij}$ and $m_{ij}$. $k_{ij}$ ($i,j=0,1$) determine how the population of tumor cells affect the immune response. $m_{ij}$ ($i,j=0,1$) quantify the death rate induced by effector cells on tumor cells. The interaction-matrix elements, $m_{ij}$ and $k_{ij}$ are not the same. For example, PD-L1+ tumor cells are able to suppress the immune response by PD-1+ immune cells, and thus the strength of $m_{11}$ can be negligible. The mechanism of adaptive immune resistance is modeled as the increase in PD-1 and PD-L1 expression levels, which can be quantified by the transition rates $p_0$ and $q_0$.
}
\label{immune1}
\end{figure}

We divide the population of tumor infiltrating T cells into low- and high- expression of PD-1 protein, namely, PD-1- and PD-1+ T cells. Similarly, we focus on two subpopulations of PD-L1+ and PD-L1- tumor cells. We denote tumor PD-L1- (+) with $y_{0}$ ($y_{1}$) and PD-1- (+) effector cells with $x_{0}$ ($x_{1}$), respectively. $y_{i}$ or $x_{i}$ for $i=0, 1$ are dynamical variables and their values can change in time depending on how immune-tumor interaction dynamics is going and how other tumor microenvironmental factors, including drug efficacies, affect them. 

To be concrete, we use the following set of biologically plausible assumptions to construct our model:

\begin{itemize}
\item The primed T-cells are supplemented at a constant rate from the lymph nodes into the tumor microenvironment. This rate is denoted with $\lambda$. 
\item The majority tumor cell population is initially PD-L1- and sensitive to the anti-tumor immune response.
\item We assume a mechanism of immune resistance where with rate, $q_{0}$, PD-L1- tumor cells transform into PD-L1+ cells. Similarly, the effector cell population transforms from PD-1- into PD-1+ with a rate $p_{0}$.
\item Any T-cell subtype (PD-1+/-) can in interact with any tumor cell subtype (PD-L1+/-). Their interaction strengths are modeled into a matrix $\{k_{ij}\}$.  Parameter $k_{ij}$ represents the stimulation/inhibition level of type $i$ effector cells by type $j$ tumor cells. For PD-1- cells $i=0$ and for PD-1+ ones $i=1$. Using similar notations, $j=0,1$ applies to PD-L1-/+ subtypes. 
\item PD-L1+ subpopulation in the tumor drives the immune resistance. Upon interaction with PD-1- immune cells a PD-L1- tumor cell stimulates the response ($k_{00} > 0$), while PD-L1+ tumor cells suppress anti-tumor response of PD-1+ T cells ($k_{11} < 0$). 
\end{itemize}

Therefore, we set $k_{00}$, $k_{01}$ and $k_{10} > 0 $ while $k_{11} < 0 $ represents the immune suppression. The anti-tumor activity of the immune cell population is modeled with a matrix of killing rate, $\{m_{ij}\}$. $m_{ij}$ is the death rate induced by a type $j $ immune cell on a type $i $ tumor cell ($i,j=\{0,1\}$). The PD-L1+ subpopulation is assumed to have lesser death rate when encountering effector cells, especially PD-1+ T cells. The above dynamics is described by the following system of equations for four subpopulations of PD-1-/+ immune cells, $x_{0}$ and $x_1$, and PD-L1-/+ tumor cells, $y_{0}$ and $y_1$:
\begin{equation}
\begin{split}
\dot{x}_{0} &= \lambda - \mu  x_{0} + \big( k_{00}y_{0}  + k_{01}y_{1}\big)x_{0} - p_{0}x_{0},\\
\dot{x}_{1} &=  		    - \mu  x_{1} + \big( k_{10}y_{0}  + k_{11}y_{1}\big)x_{1} + p_{0}x_{0},\\
\dot{y}_{0} &= ay_{0}\big( 1- b(y_{0} + y_{1})\big) - \big(m_{00}x_{0} + m_{01}x_{1}\big)y_{0} - q_0 y_0,\\
\dot{y}_{1} &= ay_{1}\big( 1- b(y_{0} + y_{1})\big) - \big(m_{10}x_{0} + m_{11}x_{1}\big)y_{1} + q_0 y_0.
\end{split}
\label{model0}
\end{equation}
Here, $\lambda$ is the recruiting rate of the effector cells from the lymph nodes to the local tumor microenvironment. $\mu$ are the death rates of PD-1+/- effector cells. $a$ is the linear growth rate of tumor and $b$ is the inverse carrying capacity for tumor cells. PD-1- T cells transform into PD-1+ with rate $p_{0}$ while PD-L1- cell become PD-L+ with rate $q_{0}$.  Values of $p_{0}$ and $q_{0}$ together quantify how fast the immune resistance is developed against an anti-tumor immune response. The above mechanisms are schematically presented in Fig.~\ref{immune1}.

We can rewrite the Eqs. \ref{model0} by rescaling tumor population with carrying capacity $b^{-1}$ and immune population with $\lambda$.

\begin{align}
b \cdot y_{0,1} &\to y_{0,1} \nonumber\\
\lambda^{-1} \cdot x_{0,1} &\to x_{0,1} 
\label{rescale1}
\end{align}

This leads to rescaling of model parameters:

\begin{align} 
b^{-1} k_{ij} \to k_{ij} \nonumber\\
\lambda^{-1} m_{ij} \to m_{ij}
\label{rescale2}
\end{align}

\nd where the subscript index $i,j \in \{0,1\} $. Eqs. \ref{model0} can be rewritten in terms of rescaled parameters as,

\begin{equation}
\begin{split}
\dot{x}_{0} &= 1 - \mu x_{0} + \big( k_{00} y_{0} + k_{01}y_{1}\big)x_{0} - p_{0}x_{0}\\  
\dot{x}_{1} &= ~-\mu x_{1} + \big( k_{10}y_{0} + k_{11}y_{1}\big) x_{1} + p_{0}x_{0}\\
\dot{y}_{0} &= a \cdot y_{0} \Big( 1- \big(y_{0} + y_{1}\big)\Big) - \big(m_{00}x_{0} + m_{01}x_1\big)y_{0} - q_{0}y_{0}\\
\dot{y}_{1} &= a \cdot y_{1} \Big( 1- \big(y_{0} + y_{1}\big)\Big) - \big(m_{10}x_{0} + m_{11}x_1\big)y_{1} + q_{0}y_{0}
\label{ode-rescaled}
\end{split}
\end{equation}

\begin{figure}
\centering
  \includegraphics[width=0.8\textwidth]{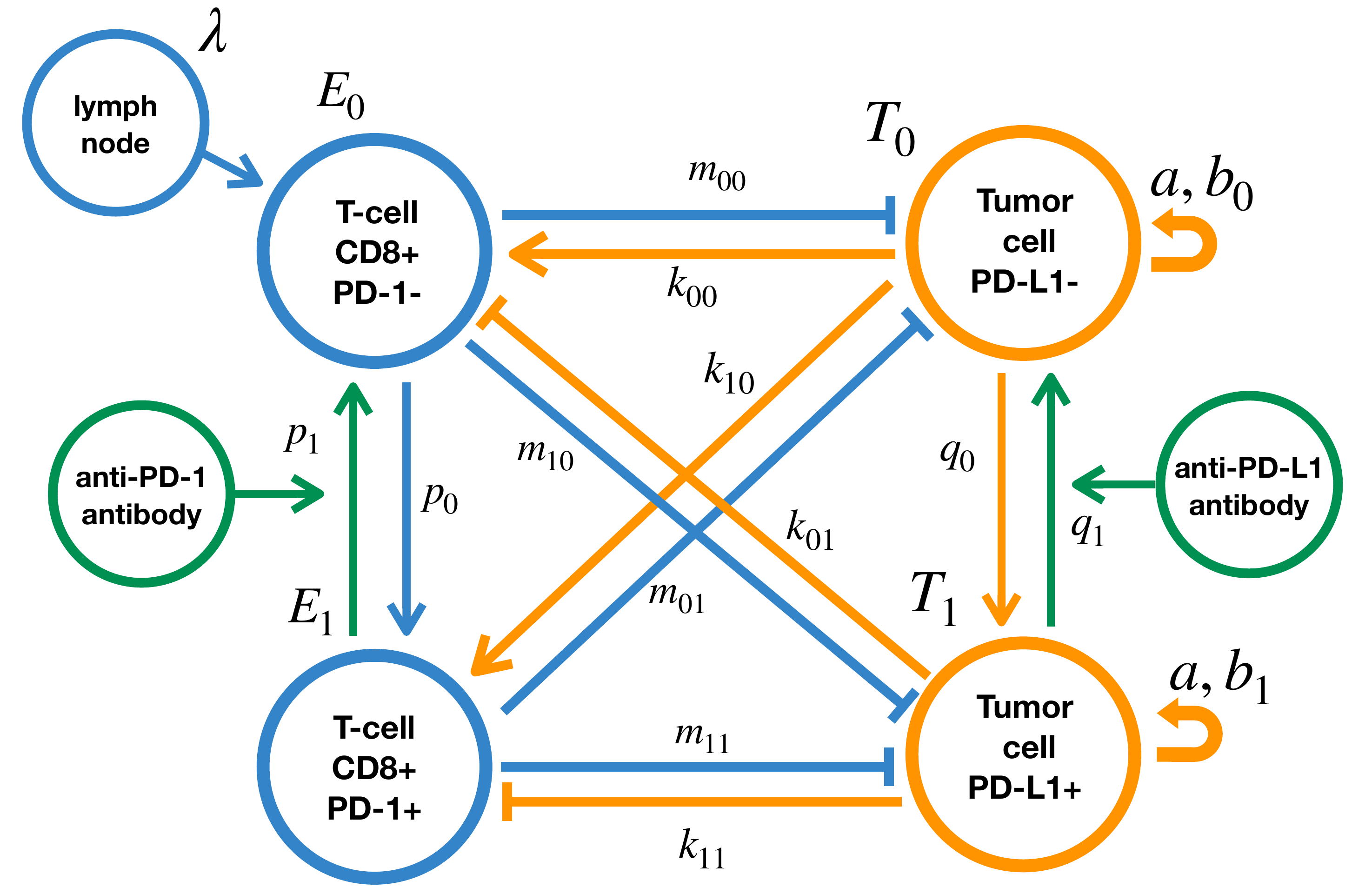}
\caption{Model scheme of tumor-immune interactions in the presence of immune checkpoint blockades. Similar to Fig.~\ref{immune1}, the effector cell population is divided into PD-1+ and PD-1- compartments and tumor cells are divided into two compartments of PD-L1+ and PD-L-. Immune-escape mechanisms cause transition from PD-L1- to PD-L1+ compartment, with rate $q_{0}$. We assume similar transition from PD-1- to PD-1+ for effector cells with rate $p_0$. Anti-PD-1 and anti-PD-L1 antibodies cause a reversal of this with corresponding rates $p_1$ and $q_1$.
}
\label{immune2}
\end{figure}

Such rescaling reduces the effective number of model parameters by two and thus facilitates our sensitivity analysis of model parameters. We could further rescale time, $t$, in the derivative, and write it in units of average life time of immune cells, $\mu^{-1}$. This would reduce the model parameters to one less. However, we keep variable $\mu$ as we want to keep the time units in specific units such as days (or proper tumor cell generation times).

We summarize the key parameters and their biological interpretations in our model as follows:

\begin{itemize}
\item[] {\bf Pre-existing immune resistance parameters, ($p_{0}$, $q_{0}$)}. These basically quantify how impaired the immune response is due to $\gamma$-interferon pathway activation and up-regulation of PD-1 on T cells and PD-L1 on tumor cells.  

\item[] {\bf Tumor growth parameters ($a$, $b$)}. We already re-scaled the model system to absorb the tumor inverse carrying capacity $b$ into other parameters. Thus the main indicator of tumor growth and aggressiveness is the value of its linear growth rate $a$.

\item[] {\bf Tumor-immune interaction matrices ($\{k_{ij}\}, \{m_{ij}\}$)} . The value of $m_{ij}$ identifies the set of death rates values for the tumor cells due to immune killing ($i, j = \{0 , 1\}$). While $k_{ij}$ determines how the strength of immune response (abundance of effector cells) is mediated by the interaction of PD-1+/- effector cells with PD-L1+/- tumor cells.

\end{itemize}

At this stage the above model can describe how quickly the tumor develops resistance to an anti-tumor immune response. Subpopulation of PD-L1- among tumor cells is basically `sensitive' to the immune response. It means that PD-L1- cells can stimulate T cells response and also are lysed after an encounter with tumor specific T cells.  PD-L1+ cells, however, have the capability to suppress the immune response. This effect in fact is not uniform and PD-L1+ cells effectively suppress immune response in PD-1+ cells and less so in PD-1- immune cells. 

The main mechanism behind the above dynamics is that in the interactions between PD-1- T cells and PD-L1- tumor cells, the tumor cells are vulnerable to the tumor-specific response from PD-1- effector cells and thus immune response is effective. In contrast, owing to the checkpoint pathway engagement via  PD-1+/PD-L1+, the immune response can be greatly impaired over time when PD-L1-  tumor cells gain the capability to switch into PD-L1+ status and thus will be able to resist the tumor-specific response and even suppress it.

\section{Results}

\subsection{Evolution of adaptive immune resistance} 

We first focus on the tumor-immune interaction dynamics in the absence of anti-PD-1 and anti-PD-L1 treatments. The dynamical model, as given in Eqs. \ref{ode-rescaled}, describes the tumor growth over time, while effector cells respond to suppress the tumor growth. The anti-tumor activity of effector cells increases upon recognizing PD-L1- tumor cells while PD-L1+ tumor cells suppress PD-1+ effector cell activity upon interaction. The transformation from PD-L1- to PD-L1+ compartments (through up-regulation of PD-L1 ligand) results in tumor immune resistance. The parameter $q_{0}$ quantifies how tumor population transforms into immune-suppressive subpopulation of PD-L1+. In the meantime, at rate $p_0$, PD-1- immune cells become PD-1+ cells with poor anti-tumor characteristic. The finite value of $q_{0}$ describes a gradual change of PD-L1- to PD-L1+, thereby indicating the level of PD-L1 expression. As a consequence of adaptive immune resistance, anti-tumor response is suppressed due to high proportion of PD-L1+ tumor cells and low abundance of PD-1- T cells (Fig.~\ref{immune2}).

\begin{figure}[h]
\centering
  \includegraphics[width=0.8\textwidth]{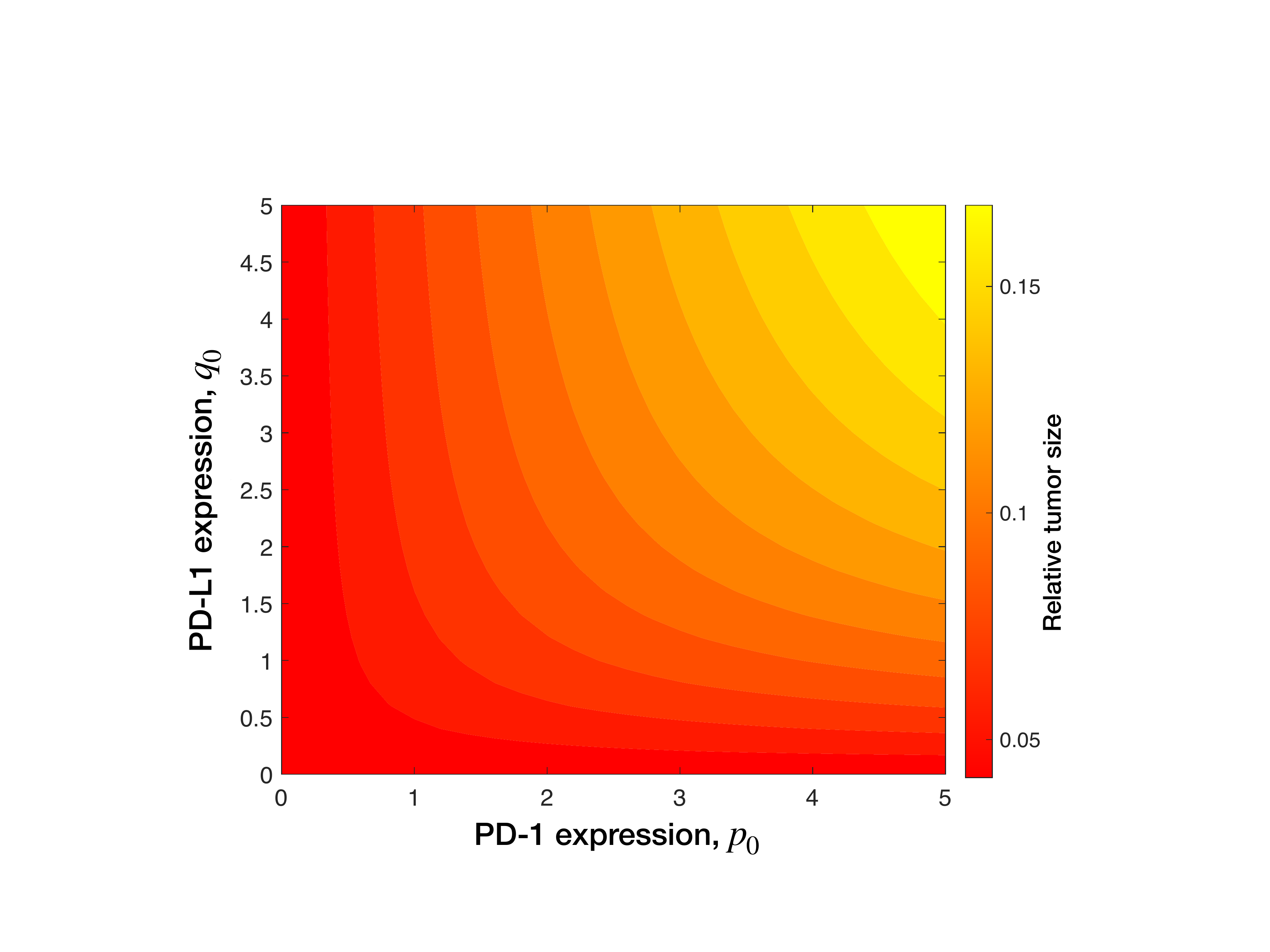}
\caption{Heatmap plot of tumor steady-state size as a function of immune resistance parameters, $p_{0}$, and $q_{0}$. We assume no treatment is applied, $q_{1}=p_{1} =0$. Tumor burden is shown as rescaled according to Eq.~\eqref{rescale1}, and relevant model parameter values are rescaled according to Eq.~\eqref{rescale2}. The immune inhibition/suppression parameters are $k_{00}=k_{01}=k_{10}=1$ and $k_{11}=-1$, and the tumor killing rates by PD-1+/- cells are $m_{00}=m_{10}=m_{01}=0.1$ and $m_{11}=0$. And we set $a=0.5$ and $\mu=0.5$ the same as in Fig.~\ref{immune2}.}
\label{immune3}
\end{figure}

Without loss of generality, we assume the values of immune stimulation/inhibition, $k_{00}, k_{01}, k_{10}$ are the same order of magnitude. We model suppressive effect of PD-L1+ tumor cells on immune cells by setting $k_{11}$ to a negative value. Anti-tumor activity of effector cells is parametrized by prescribing values of $m_{ij}$. In the example shown in Fig.~\ref{immune2}, we use $m_{00}=m_{01}=m_{10}=0.1$ and $m_{11}=0$. The value of $m_{11}$ indicates that PD-1+ T-cells has no effector activity on PD-L1+ tumor cells. The tumor-immune interaction parameter, $k_{ij}$ are set to $k_{00}=k_{10}=k_{01}=1$, and $k_{11}= - 1$. The positive (negative) values of $k_{ij}$ mean stimulatory (or inhibitory) response due to immune-tumor interaction, respectively. For the rest of parameters we used $a=0.5, \mu=0.5$ and values of $p_{0}=q_{0}=1,2,3$. In Fig.~\ref{immune2}, we plot the total effector cell count, $x_{0} + x_{1}$, and the total tumor size, $y_{0} + y_{1}$ as a function of time, using the initial condition $x_{1}(0) = y_{1}(0) = 0$. 
The higher the values of $p_{0}$ and $q_{0}$ are, the dynamics is closer to free tumor growth as if in the absence of immune-tumor interactions. The equilibrium tumor size (obtained when the above model reaches the steady state) as a function of immune resistance parameters $(p_0, q_0)$ is shown in Fig.~\ref{immune3}. Same model parameters as in Fig.~\ref{immune2} are used, and values of $p_{0}$ and $q_{0}$ are varied independently. As adaptive immune resistance is modulated by the $\gamma$-interferon pathway activation and up-regulation of PD-1 on T cells and PD-L1 on tumor cells, complete immune escape occurs only for high levels of PD-1 and PD-L1 expression, that is, large values of $p_0$ and $q_0$ (see Fig.~\ref{immune3}).


To obtain closed-form results beside numerical simulations, we can simplify the immune-tumor interaction matrix by writing four components in terms of a single parameter $\beta$: $k_{00}=k_{01}=k_{10}=-k_{11}=\beta$. A simple analytical expression can be found for the steady-state tumor size along the diagonal $q_{0}=p_{0} = \eta$: 

\begin{align}
T^{\star}_{0} + T^{\star}_{1} = y_0^* + y_1^* = \frac{a \eta- \eta^2 - \lambda \mu + a - \eta}{(a - \eta)\beta}
\label{analytic0}
\end{align}


For small-$\eta$, the formula above can be expressed in a Taylor expansion:

\begin{align}
T^{\star} &\approx  \frac{a\beta + a + \sqrt{a^2(\beta - 1)^2 + 4\beta\lambda\mu}}{2a\beta}  \nonumber\\
&+ \frac{1}{2} \frac{(-a\beta + \sqrt{a^2(\beta-1)^2 + 4a\beta\lambda\mu} + a)}{\beta\sqrt{a((\beta - 1)^2 a + 4\beta\lambda\mu)}} \eta + \mathcal{O}(\eta^2)
\end{align}

Notably, this simplified formula provides us a clear and intuitive picture about the relationship between tumor burden and levels of adaptive immune resistance. The equilibrium tumor burden $T^*$ monotonically increases with the adaptive immune resistance parameters $p_0 =q_0 = \eta$ as the coefficient of the first order expansion in $\eta$ is positive. High expression levels of PD-1 and PD-L1 lead to severely impaired immune response and as a consequence, yield high tumor burden (as shown in Fig.~\ref{immune2}).

\begin{figure}
\centering
  \includegraphics[width=0.8\textwidth]{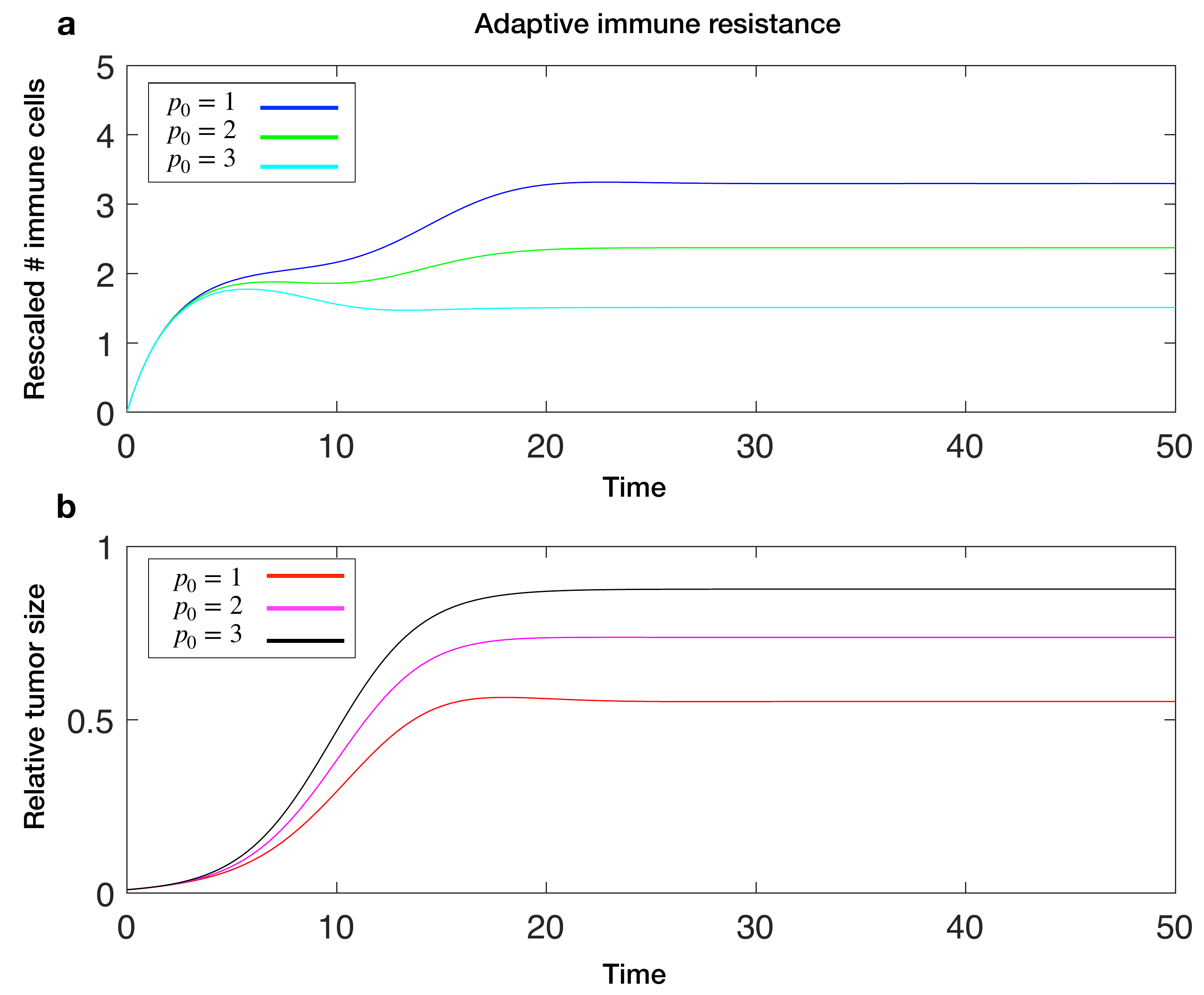}
\caption{The impact of adaptive immune resistance on tumor burden and immune cell counts. The top panel (a) shows the total number of effector cells (tumor infiltrating cells) as a sum of PD-1+ and PD-1- cells. The bottom panel (b) shows the tumor population (total of PD-L1+ and PD-L1-  cells). For different plots we have changed the value of $p_{0}=1, 2, 3$ while the value of $q_{0}$ is kept the same as $p_{0}$, $q_{0} = p_{0}$. The abundance of effector cells and tumor burden are shown as rescaled according to Eq.~\eqref{rescale1}, and relevant model parameter values are rescaled according to Eq.~\eqref{rescale2}.
Rescaled model parameters are $k_{00}=k_{01}=k_{10}=1$, $k_{11}=-1$, $m_{00}=m_{10}=m_{01}=0.1$, $m_{11}=0$, and $a =0.5$, $\mu=0.5$. 
}
\label{immune2}
\end{figure}

\subsection{Anti-PD-1 versus anti-PD-L1 monotherapy} 

We extend the above model of adaptive immune resistance by incorporating immune checkpoint therapy using PD-1 and PD-L1 blockades. Put simply, the effect of immune checkpoint blockade treatments is exerted on the PD-1+ (PD-L1+) subpopulation 
 by neutralizing and blocking PD-1 (and PD-L1) receptors and further reversing them into PD-1- (PD-L1-) subpopulation, respectively. Such conversion happens after immune cells (tumor cells) are bound with anti-PD-1 (anti-PD-L1) antibodies and thus the immune checkpoint pathway is successfully blocked. In Eqs. \ref{ode-rescaled} developing adaptive immune resistance is modeled through transition rates $p_{0}$ and $q_{0}$, which determine how quickly PD-1 ( PD-L1) expression levels of cells transition from lower (-) to higher values (+). As such, we account for the effect of immune checkpoint therapy by corresponding reverse transition rates. That is, in the presence of anti-PD-1 antibodies, PD-1+ immune cells transition to PD-1- compartment with the rate $p_{1}$. The bulk parameter $p_{1}$ reflects the overall neutralization (blockade) efficacy determined by the underlying concentration-dependent binding kinetics~\cite{brown2020assessing}. Similarly, we denote by $q_{1}$ the reverse transition rate for tumor cells transitioning from PD-L1+ to PD-L1- compartments in the presence of anti-PD-L1 treatment.

%

Incorporating the anti-PD-1/anti-PD-L1 treatment into the Eqs.~\ref{ode-rescaled}, we obtain the following system of differential equations:
\begin{equation}
\begin{split}
\dot{x}_{0} &= 1 - \mu x_{0} + \big( k_{00} y_{0} + k_{01}y_{1}\big)x_{0} - p_{0}x_{0} + p_{1}x_{1},\\  
\dot{x}_{1} &= ~-\mu x_{1} + \big( k_{10}y_{0} + k_{11}y_{1}\big) x_{1} + p_{0}x_{0} - p_{1}x_{1},\\
\dot{y}_{0} &= a \cdot y_{0} \Big( 1- \big(y_{0} + y_{1}\big)\Big) - \big(m_{00}x_{0} + m_{01}x_1\big)y_{0} - q_{0}y_{0} + q_{1}y_{1},\\
\dot{y}_{1} &= a \cdot y_{1} \Big( 1- \big(y_{0} + y_{1}\big)\Big) - \big(m_{10}x_{0} + m_{11}x_1\big)y_{1} + q_{0}y_{0} - q_{1}y_{1}.
\end{split}
\label{model1}
\end{equation}

Closed-form analytical solutions can be obtained for the steady-state tumor size, $T^{\star} = T^{\star}_{0} + T^{\star}_{1}$ (that is, $y_0^* + y_1^*$), if we apply the same assumption as in Eq. \ref{analytic0}, $k_{00}=k_{01}=k_{10}=-k_{11}=\beta, q_{0}=p_{0} = \eta$, as well as $p_{1}=\zeta, q_{1}=0$ for anti-PD-1 treatment. We have,

\begin{align}
T^{\star} = \frac{(\eta - \zeta)a - \lambda\mu + \sqrt{a^2\zeta^2 - 2a(a\eta + \lambda\mu - 2a)\zeta + (a\eta - \lambda\mu + 2a)^2}}{2a\beta}.
\end{align}

Similar results can be obtained in the case of PD-L1 treatment.

It is worthy noting that while treatment efficacy is quantified mainly by model parameters $p_1$ and $q_1$ in this paper, the parameter $k_{11}$ characterizing the tumor-immune interaction via the checkpoint pathway PD-1+/PD-L1+ can be affected to some extent during treatment. PD-1+ T cells are the primary target of anti-PD-1 therapy, and after PD-1 blockade, they effectively become PD-1- cells that are responsive again to tumor-immune stimulatory signals. As PD-1 blockade antibody can also competitively weaken the binding activities between PD-1+/PD-L1+, we assume anti-PD-1 treatment is able to reduce the magnitude of immune suppression signaling $k_{11}$. Similarly, we assume anti-PD-L1 therapy targets PD-L1+ tumor cells, and meanwhile can also reduce the value of tumor-immune interaction matrix $k_{11}$. Hence, the parameter value of $k_{11}$ during treatment is uniformly altered as a secondary effect of monotherapy or combination therapy. 

\begin{figure}
\centering
  \includegraphics[width=0.6\textwidth]{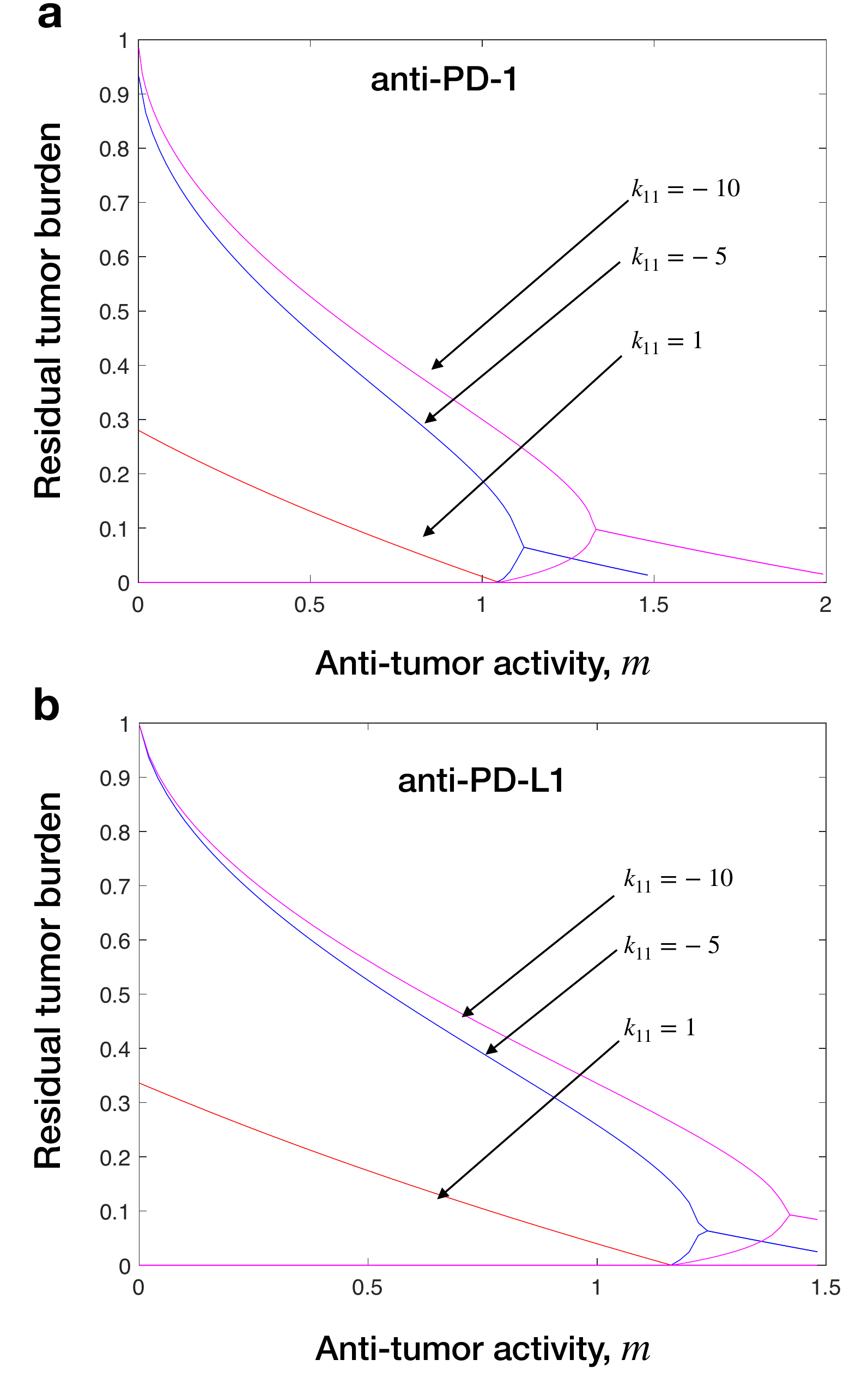}
\caption{Bifurcation analysis of equilibrium tumor burden with respect to immune killing rate $m$ under monotherapy with anti-PD-1 and anti-PD-L1. tumor fixed point, $T^{\star} = T^{\star}_{0} + T^{\star}_{1}$, for various values of $m$ (tumor killing rate) for PD-L1 mono-therapy as immune-tumor interaction parameter $k_{11}$ is varied due to treatment. Tumor burden is shown as rescaled according to Eq.~\eqref{rescale1}, and relevant model parameter values are rescaled according to Eq.~\eqref{rescale2}. Model parameters are set as $m_{00}=m_{01}=m_{10}=m$, $m_{11}=0.01m$, $q_{0}=p_{0}=5$, $k_{00}=k_{10}=k_{01}=5$, $k_{11} = -10, -5, 1$, (a) anti-PD-1: $p_{1}=1, q_{1}=0$, and (b) anti-PD-L1: $p_{1}=0, q_{1}=1$. And we set $a=0.5$ and $\mu=0.5$ the same as in Fig.~\ref{immune2}.
}
\label{bifur}
\end{figure}

We perform bifurcation analysis of equilibrium tumor burden, $T^*$, with respect to changes in the anti-tumor immune activity, $m_{ij}$. In Fig.~\ref{bifur}, we plot the steady-state solutions (fixed points) for anti-PD-1 and anti-PD-L1 treatments, respectively. We also vary values of $k_{11}=+1, -5, -10$ while other immune-tumor interaction parameters are kept constant $k_{00}=k_{01}=k_{10}=5$. Values of tumor killing rates $m_{ij}$ are parametrized as $m_{00}=m_{01}=m_{10}=m, m_{11}=0.01m$. In other words, the PD-1+/PD-L1+ interaction leads to much weaker cytotoxic activity from T-cells. For both anti-PD-1 and anti-PD-L1 monotherapies,  we observe an interesting \emph{bistability} behavior for $k_{11} < 0$ and for intermediate levels of anti-tumor activity (Fig.~\ref{bifur}). One potential implication of bistablity phenomena is that the existence of an interior unstable equilibrium allows the occurrence of hysteresis. Namely, the rebound of tumor mass after treatment stops has to follow the lower branch instead of the upper branch along which it plummeted during treatment, and thus tumor can be controlled at low abundance until the anti-tumor activity is decayed below a certain threshold. 

%
%
%

\begin{figure}
\centering
  \includegraphics[width=0.8\textwidth]{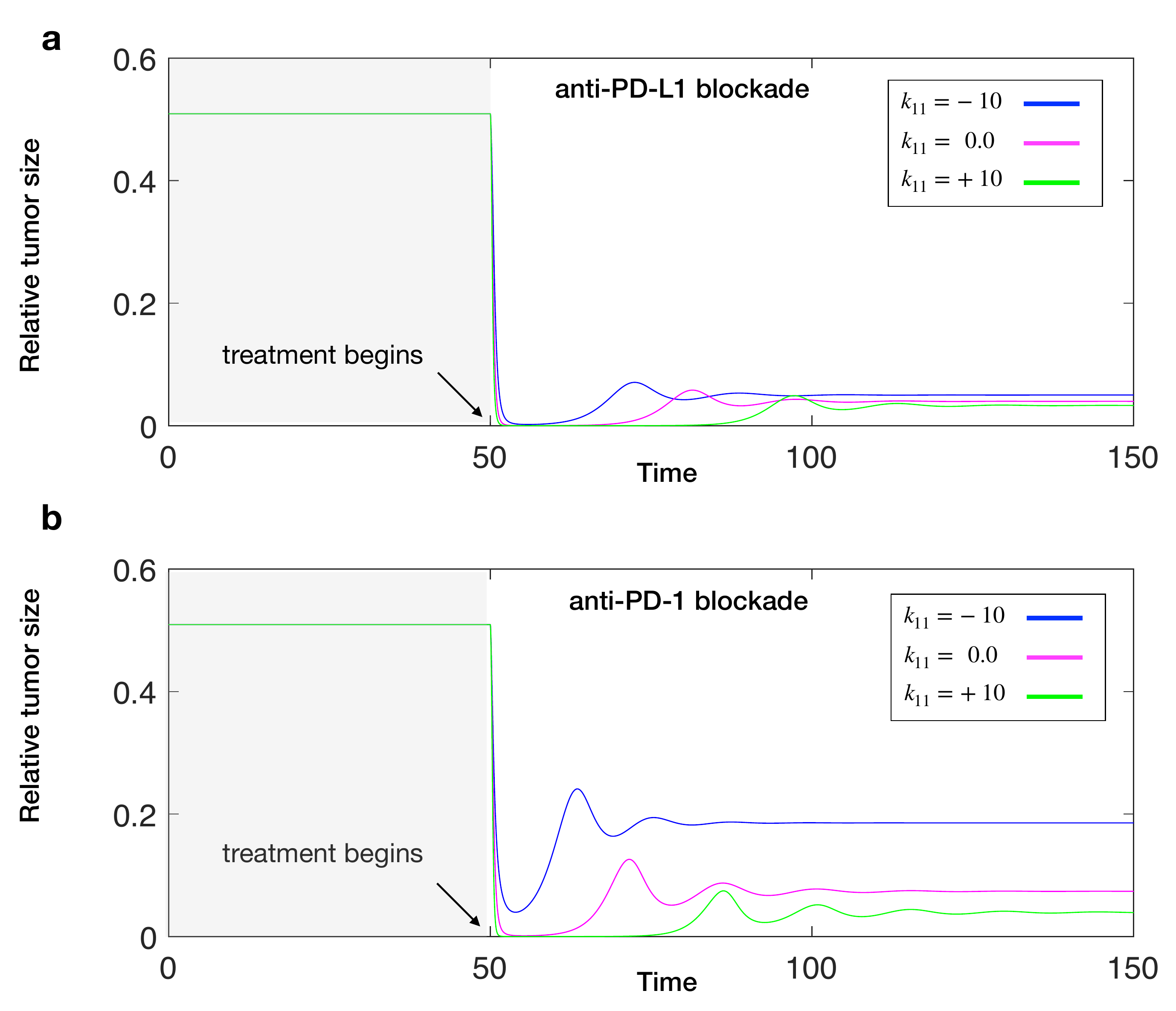}
\caption{Comparison of anti-PD-L1 (upper panel) and anti-PD-1 (lower panel) monotherapy (lower panel). The relative tumor size (to carrying capacity) is used as a quantitative measure of response dynamics and treatment outcomes. Treatment begins with tumor at its steady state in the absence of immune checkpoint therapy. In this example, anti-PD-L1 monotherapy seems to be more effective (i.e., yielding smaller residual tumor burden) than anti-PD-1 monotherapy, because of higher expression level of PD-1 than that of PD-L1 ($p_{0}=5,q_{0}=1$). Tumor burden is shown as rescaled according to Eq.~\eqref{rescale1}, and relevant model parameter values are rescaled according to Eq.~\eqref{rescale2}. Other model parameters used are: $k_{00}=k_{10}=k_{10}=10$, $k_{11}=-10, 0,10$, $m_{00}=m_{01}=m_{10}=0.1$, $m_{11}=0$, (a) anti-PD-L1: $p_{1}=0, q_{1}=5$, and (b) anti-PD-1: $p_{1}=5,q_{1}=0$. And we set $a=0.5$ and $\mu=0.5$ the same as in Fig.~\ref{immune2}.}
\label{immune4}
\end{figure}

\begin{figure}
\centering
  \includegraphics[width=0.5\textwidth]{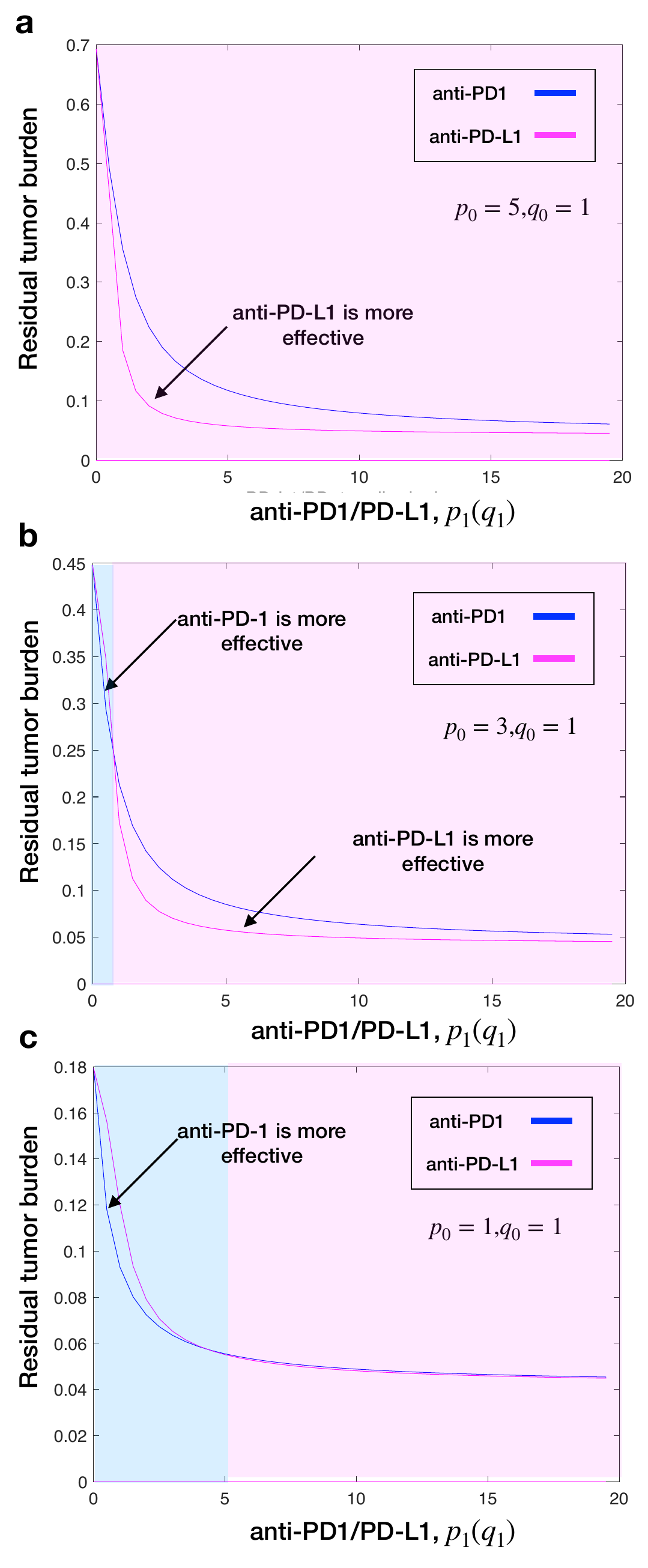}
\caption{Residual tumor burden as a function of treatment efficacy of monotherapy, anti-PD-1 $p_1$ versus anti-PD-L1 $q_1$. In panels (a)-(c), anti-PD-1 and anti-PD-L1 monotherapies are compared by assuming the same degree of efficacy expressed in terms of neutralizing PD-1+ and PD-L1+ cells: for anti-PD-L1 we have $q_{1}=p, p_{1}=0$ for anti-PD-1 we have $q_{1}=0,p_{1}=p$, where $p$ denotes the x-axis. Depending on the expression levels of PD-1 and PD-L1 status (immune resistance parameters $p_0$ and $q_0$), there exists an interesting crossover of the two tumor burden curves as shown in panels (b) and (c): anti-PD-1 is slightly more effective for low neutralizing efficacies, but  anti-PD-L1 seems to work better for high neutralizing efficacies. Rescaled model parameters are, $m_{00}=m_{01}=m_{10}=0.1$, $m_{00}=0.01$, $k_{00}=k_{01}=k_{10}=7$, $k_{11}=-1$, $a=0.5$, $\mu=0.5$, (a) $p_0 = 5, q_0 = 1$, (b) $p_0 = 3, q_0 = 1$, (c) $p_0 = 1, q_0 = 1$.
}
\label{immune5}
\end{figure}

To further reveal the subtle difference between anti-PD-1 and anti-PD-L1 monotherapies, we plot the time evolution of tumor burden during treatment in Fig. \ref{immune4}. For anti-PD-1 monotherapy, we set $q_{1}=0$ while $p_{1} > 0$. Similarly, for anti-PD-L1 monotherapy, $q_{1} > 0 $ while $p_{1} = 0$. To compare the efficacy of the two treatments we keep the remaining model parameters exactly the same. In the examples shown in Fig. \ref{immune4}, we note that the treatment outcome of anti-PD-L1 therapy is better than anti-PD-1 therapy for the specific parameter choices ($p_0 = 5$ and $q_0=1$) that indicate high PD-1 expressions prior to treatment (also see Fig.~\ref{immune5}a). More generally, these exists an interesting crossover of the residual tumor curves (as shown in Figs.~\ref{immune5}b and ~\ref{immune5}c): anti-PD-1 is slightly more effective for low neutralizing efficacies, but anti-PD-L1 works better for high neutralizing efficacies. These theoretical results suggest that PD-L1 expression level alone is insufficient to determine which anti-PD-1 or anti-PD-L1 blockade to be more effective~\cite{shen2018efficacy}. However, for a wide range of model parameters, it appears that anti-PD-L1 therapy is more effective than anti-PD-1 therapy even for equally potent PD-1 and PD-L1 blockades~\cite{linhares2019therapeutic}.

\begin{figure}
\centering
  \includegraphics[width=0.8\textwidth]{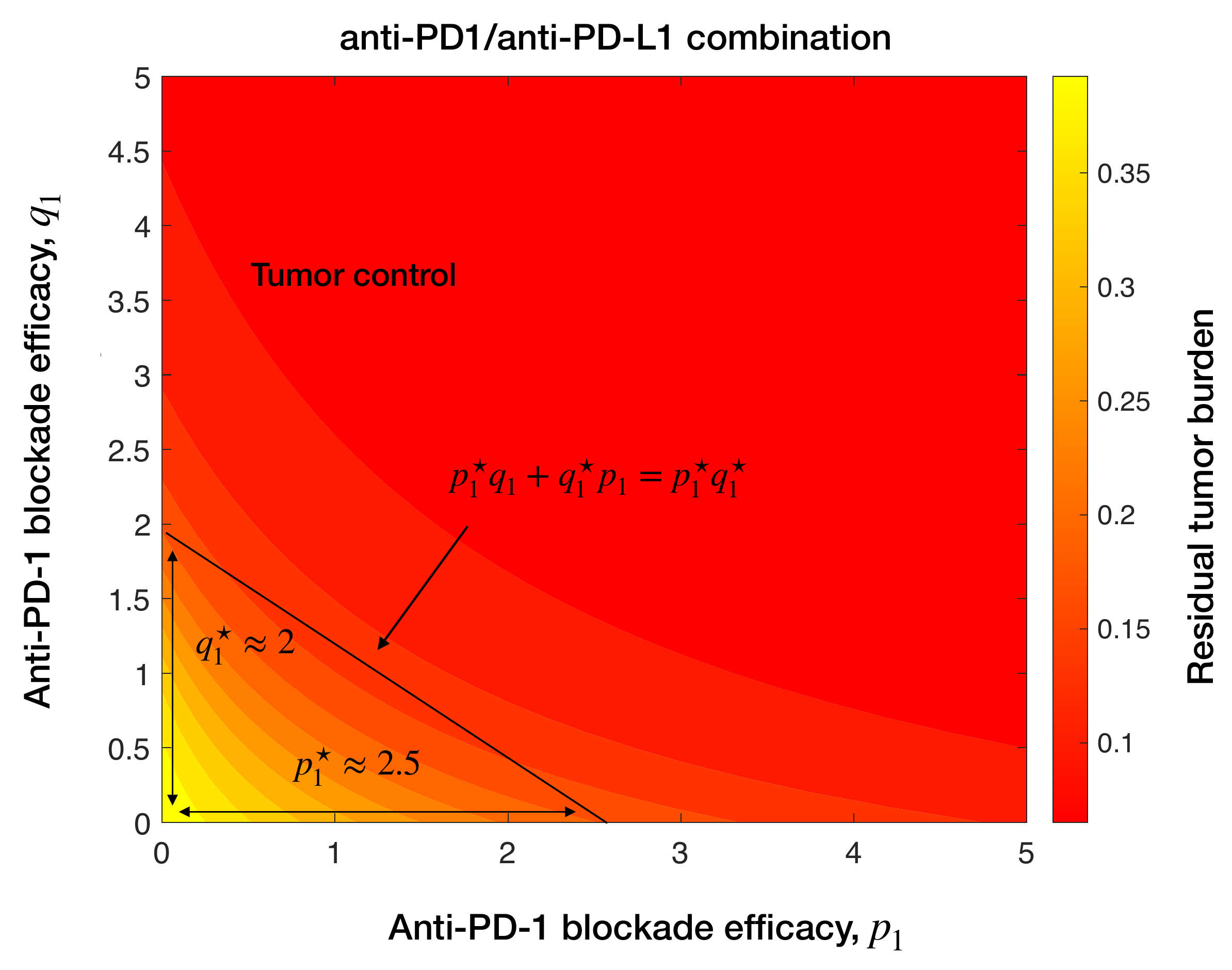}
\caption{Synergistic effect of anti-PD-L1 and anti-PD-1 combination immunotherapy. Residual tumor burden is shown as a contour plot of the parameter space $(p_1, q_1)$, which quantifies anti-PD-1 and anti-PD-L1 treatment efficacies. The Loewe additivity is indicated by the straight line $p^{\star}_1q_1 + p_1q^{\star}_1 = p^{\star}_1q^{\star}_1$, where $q^{\star}_1$ and $p^{\star}_1$ are the critical values for each corresponding monotherapy to have the same treatment effect (namely, residual tumor burden). The contour lines are convex downward, suggesting synergy between anti-PD-L1 and anti-PD-1 treatments. Tumor burden is shown as rescaled according to Eq.~\eqref{rescale1}, and relevant model parameter values are rescaled according to Eq.~\eqref{rescale2}. Model parameters are $\mu=0.5,m_{00}=m_{01}=m_{10}=0.1,m_{11} = 0.01, k_{00}=k_{01}=k_{10}=5,k_{11}=-5,a=0.5,p_0=2,q_0=1$.}
\label{immune6}
\end{figure}


\begin{figure}
\centering
  \includegraphics[width=0.8\textwidth]{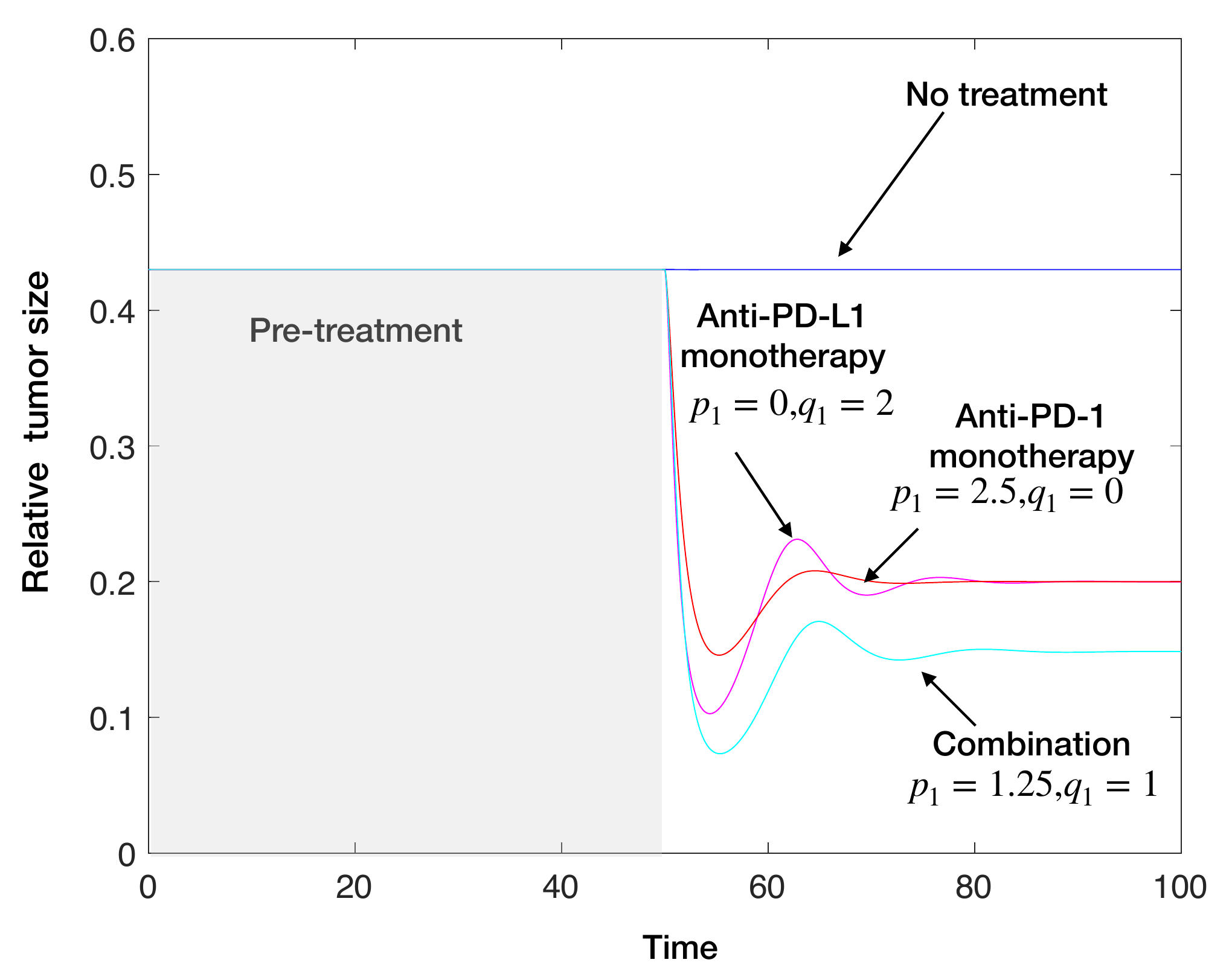}
\caption{Immunotherapy combining PD-1 and PD-L1 blockades renders better tumor shrinkage than monotherapy. For the sake of proper comparison in accordance with the Loewe synergy~\cite{loewe1926kombinationswirkungen}, we restrict parameter choices of $p_1$ and $q_1$ for combination therapy along the line (see Fig. \ref{immune6}), $p^{\star}_1q_1 + p_1q^{\star}_1 = p^{\star}_1q^{\star}_1$, where $q^{\star}_1$ and $p^{\star}_1$ are the critical values for each corresponding monotherapy to have the same treatment effect (namely, residual tumor burden). We assume tumor reaches steady state prior to treatment, and the relative tumor size is plotted as a function of time after the treatment starts at $t=50$. Monotherapy: anti-PD-1 $p_1 = 2.5, q_1 = 0$, anti-PD-L1 $p_1 = 0, q_1 = 2$, combination therapy: $p_1 = 1.25, q_1 = 1$, other rescaled model parameters: $m_{00}=m_{01}=m_{10}=0.1$, $m_{00}=0.01$, $k_{00}=k_{01}=k_{10}=5$, $k_{11}=-5$, $p_0=2$, $q_0=1$, $a=0.5$, $\mu=0.5$.}
\label{immune7}
\end{figure}


\subsection{Anti-PD-1 and anti-PD-L1 combination immunotherapy}

It is straightforward to investigate combination immunotherapy based on Eqs.~\eqref{model1}. Immune checkpoint therapy combining anti-PD-1 and anti-PD-L1 blockades is charaterized by the parameter space $(p_1, q_1)$, which quantifies the treatment efficacy of each checkpoint blockade alone. A natural question is whether there exists Loewe synergy between anti-PD-1 and anti-PD-L1 treatments~\cite{loewe1926kombinationswirkungen}.

In Fig.~\ref{immune6}, we show a contour plot of residual tumor burden as a function of $(p_1, q_1)$. Along each contour line, the residual tumor burden is constant for the combination of values of  $(p_1, q_1)$. Let us denote by $p_1^*$ ($q_1^*$) the critical efficacy of monotherapy which is needed to reach a fixed value of residual tumor burden. If anti-PD-1 and anti-PD-L1 treatments satisfy the Loewe additivity,  that is, for any combination of $p_1$ and $q_1$ satisfying
\begin{equation}
q_1^* p_1 + p_1^*q_1 = p_1^* q_1^*.
\end{equation}
we have the exactly same treatment effect where residual tumor burden always reaches the same value. However, the contour lines in Fig.~\ref{immune6} are not straight but instead convex downward, which suggest synergy between anti-PD-1 and anti-PD-L1 blockades. Figure~\ref{immune7} further demonstrates that neither of monotherapies is able to yield greater tumor shrinkage than combination therapy.

\section{Discussion \& Conclusion}

In recent years, there has been a surge of interest in developing immune-checkpoint inhibitors to treat cancer~\cite{tang2018comprehensive}. While these potentially curative cancer therapies are rapidly being developed and tested, a major barrier is the lack of quantitative models of their efficacy. To address this issue, mathematical modeling of cancer-immune interactions and of immunotherapy has been a topic of interest and primary significance~\cite{depillis2014modeling,de2005validated,wilson2012mathematical,yamamoto2016tumour,serre2016mathematical,castiglione2007cancer,owens2020modelling,kimmel2020response}. Prior mathematical models of cancer-immune interactions~\cite{kirschner1998modeling,eftimie2011interactions} are mostly based on ordinary differential equations in combination with stochastic \emph{in-silico} simulations~\cite{lakatos2020evolutionary,west2019immune}. Building on and integrating these aforementioned models, here we focus on quantifying and predicting the efficacy of  immune checkpoint PD-1/PL1 blockades treatment and their combinations.

An open question in the field is, other things being equal, which monotherapy, anti-PD-1 or anti-PD-L1 is more effective in tumor control and eradication provided they have the same checkpoint blockade binding efficacy~\cite{linhares2019therapeutic,chen2018sequential}. Our results show that depending on PD-1 and PD-L1 expression levels, anti-PD-1 or anti-PD-L1 treatment can be more effective than the counterpart. However, in most cases anti-PD-L1 treatment is seen to be a more effective treatment in a wide range of model parameters. Altogether, our work demonstrates that PD-L1 expression level alone is insufficient to determine which anti-PD-1 or anti-PD-L1 blockade to be more effective~\cite{shen2018efficacy}.

Randomized trials of monotherapy and combination therapy for treating advanced cancers are underway and some have been completed~\cite{sui2018anti,sato2020rationale}. Our theoretical modeling predicts synergy between anti-PD-1 and anti-PD-L1 treatments. This prediction is in line with a recent work, which shows that PD-1/PD-L1 blockade combination leads to better treatment outcomes among metastatic breast cancer patients~\cite{page2019two}. Future work can  incorporate into the current models with cancer and immune biomarkers measured before and during immunotherapy~\cite{willis2015immune}, and develop theory-informed combination therapies to overcome immunotherapy resistance and improve response.

In this work, the switching rate of effector cells, $p_{0}$ (and similarly tumor cells, $q_{0}$) to increase the expression level of PD-1  (PD-L1 ligand, respectively) is assumed to be constant as a proof-of-principle model with a parsimonious number of parameters. It is promising for future work to incorporate detailed molecular reaction kinetics and dynamics associated with the PD-1/PD-L1 regulation pathway~\cite{wang2019silico,milberg2019qsp} and also spatial infiltration of T cells into tumor mass~\cite{gong2017computational}. These meaningful extensions may further improve our understanding of the subtle differences in treatment outcomes between anti-PD-1 and anti-PD-L1 immune checkpoint blockades.


In summary, we have proposed a proof-of-concept, dynamical systems model of tumor-immune interactions along the PD-1/PD-L1 axis. Based on this model, we have studied the evolution of adaptive immune resistance and subsequent response to anti-PD-1/PD-L1 treatments. Depending on model parameters describing immune resistance, tumor growth and treatment efficacies, our modeling framework provides mechanistic insights into quantifying and characterizing conditions for the success (complete or partial response) or failure of immunotherapy using checkpoint inhibitors and their potential combinations. Further parameterized with patient-specific tumor and immune biomarker data~\cite{brady2019mathematical}, our models can be used to test hypothetical drug administration schedules \emph{in-silico} and to optimize cancer treatment in a personalized fashion.

%
%
%

\section*{Author contributions}
K.K. \& F.F. conceived the model and performed theoretical analysis; K.K. conducted numerical simulations, plotted figures, and wrote the first version of the draft; F.F. secured funding and supervised the project; K.K. \& F.F. contributed to the revision of the draft and gave approval of the final manuscript.

\section*{Acknowledgments}
This work is supported by the NIH COBRE Program (grant no. 1P20GM130454). F.F. is grateful for the generous financial support by the Bill \& Melinda Gates Foundation (award no. OPP1217336) and the Neukom CompX Faculty Grant.



\begin{thebibliography}{10}
\expandafter\ifx\csname url\endcsname\relax
  \def\url#1{\texttt{#1}}\fi
\expandafter\ifx\csname urlprefix\endcsname\relax\def\urlprefix{URL }\fi
\expandafter\ifx\csname href\endcsname\relax
  \def\href#1#2{#2} \def\path#1{#1}\fi

\bibitem{ribas2015releasing}
A.~Ribas, et~al., Releasing the brakes on cancer immunotherapy, N Engl J Med
  373~(16) (2015) 1490--1492.

\bibitem{ribas2018cancer}
A.~Ribas, J.~D. Wolchok, Cancer immunotherapy using checkpoint blockade,
  Science 359~(6382) (2018) 1350--1355.

\bibitem{sharma2015immune}
P.~Sharma, J.~P. Allison, Immune checkpoint targeting in cancer therapy: toward
  combination strategies with curative potential, Cell 161~(2) (2015) 205--214.

\bibitem{kucerova2016spontaneous}
P.~Kucerova, M.~Cervinkova, Spontaneous regression of tumour and the role of
  microbial infection--possibilities for cancer treatment, Anti-Cancer Drugs
  27~(4) (2016) 269.

\bibitem{ribas2015adaptive}
A.~Ribas, Adaptive immune resistance: how cancer protects from immune attack,
  Cancer Discovery 5~(9) (2015) 915--919.

\bibitem{kalbasi2019tumour}
A.~Kalbasi, A.~Ribas, Tumour-intrinsic resistance to immune checkpoint
  blockade, Nature Reviews Immunology (2019) 1--15.

\bibitem{wei2018fundamental}
S.~C. Wei, C.~R. Duffy, J.~P. Allison, Fundamental mechanisms of immune
  checkpoint blockade therapy, Cancer Discovery 8~(9) (2018) 1069--1086.

\bibitem{littman2015releasing}
D.~R. Littman, Releasing the brakes on cancer immunotherapy, Cell 162~(6)
  (2015) 1186--1190.

\bibitem{tang2018comprehensive}
J.~Tang, A.~Shalabi, V.~Hubbard-Lucey, Comprehensive analysis of the clinical
  immuno-oncology landscape, Annals of Oncology 29~(1) (2018) 84--91.

\bibitem{sharma2015future}
P.~Sharma, J.~P. Allison, The future of immune checkpoint therapy, Science
  348~(6230) (2015) 56--61.

\bibitem{wei2017distinct}
S.~C. Wei, J.~H. Levine, A.~P. Cogdill, Y.~Zhao, N.-A.~A. Anang, M.~C. Andrews,
  P.~Sharma, J.~Wang, J.~A. Wargo, D.~Pe?er, et~al., Distinct cellular
  mechanisms underlie anti-ctla-4 and anti-pd-1 checkpoint blockade, Cell
  170~(6) (2017) 1120--1133.

\bibitem{wei2019combination}
S.~C. Wei, N.-A.~A. Anang, R.~Sharma, M.~C. Andrews, A.~Reuben, J.~H. Levine,
  A.~P. Cogdill, J.~J. Mancuso, J.~A. Wargo, D.~Pe?er, et~al., Combination
  anti--ctla-4 plus anti--pd-1 checkpoint blockade utilizes cellular mechanisms
  partially distinct from monotherapies, Proceedings of the National Academy of
  Sciences 116~(45) (2019) 22699--22709.

\bibitem{willis2015immune}
J.~C. Willis, G.~M. Lord, Immune biomarkers: the promises and pitfalls of
  personalized medicine, Nature Reviews Immunology 15~(5) (2015) 323--329.

\bibitem{chaplin2010overview}
D.~D. Chaplin, Overview of the immune response, Journal of Allergy and Clinical
  Immunology 125~(2) (2010) S3--S23.

\bibitem{nicholson2016immune}
L.~B. Nicholson, The immune system, Essays in Biochemistry 60~(3) (2016)
  275--301.

\bibitem{gibbons2017functional}
R.~M. Gibbons~Johnson, H.~Dong, Functional expression of programmed
  death-ligand 1 (b7-h1) by immune cells and tumor cells, Frontiers in
  Immunology 8 (2017) 961.

\bibitem{wu2019pd}
Y.~Wu, W.~Chen, Z.~P.~G. Xu, W.~Gu, Pd-l1 distribution and perspective for
  cancer immunotherapy--blockade, knockdown, or inhibition, Frontiers in
  Immunology 10 (2019) 2022.

\bibitem{parsa2007loss}
A.~T. Parsa, J.~S. Waldron, A.~Panner, C.~A. Crane, I.~F. Parney, J.~J. Barry,
  K.~E. Cachola, J.~C. Murray, T.~Tihan, M.~C. Jensen, et~al., Loss of tumor
  suppressor pten function increases b7-h1 expression and immunoresistance in
  glioma, Nature Medicine 13~(1) (2007) 84.

\bibitem{akbay2013activation}
E.~A. Akbay, S.~Koyama, J.~Carretero, A.~Altabef, J.~H. Tchaicha, C.~L.
  Christensen, O.~R. Mikse, A.~D. Cherniack, E.~M. Beauchamp, T.~J. Pugh,
  et~al., Activation of the pd-1 pathway contributes to immune escape in
  egfr-driven lung tumors, Cancer Discovery 3~(12) (2013) 1355--1363.

\bibitem{atefi2014effects}
M.~Atefi, E.~Avramis, A.~Lassen, D.~J. Wong, L.~Robert, D.~Foulad,
  M.~Cerniglia, B.~Titz, T.~Chodon, T.~G. Graeber, et~al., Effects of mapk and
  pi3k pathways on pd-l1 expression in melanoma, Clinical Cancer Research
  20~(13) (2014) 3446--3457.

\bibitem{taube2012colocalization}
J.~M. Taube, R.~A. Anders, G.~D. Young, H.~Xu, R.~Sharma, T.~L. McMiller,
  S.~Chen, A.~P. Klein, D.~M. Pardoll, S.~L. Topalian, et~al., Colocalization
  of inflammatory response with b7-h1 expression in human melanocytic lesions
  supports an adaptive resistance mechanism of immune escape, Science
  Translational Medicine 4~(127) (2012) 127ra37--127ra37.

\bibitem{tumeh2014pd}
P.~C. Tumeh, C.~L. Harview, J.~H. Yearley, I.~P. Shintaku, E.~J. Taylor,
  L.~Robert, B.~Chmielowski, M.~Spasic, G.~Henry, V.~Ciobanu, et~al., Pd-1
  blockade induces responses by inhibiting adaptive immune resistance, Nature
  515~(7528) (2014) 568.

\bibitem{pentcheva2014cytotoxic}
T.~Pentcheva-Hoang, T.~R. Simpson, W.~Montalvo-Ortiz, J.~P. Allison, Cytotoxic
  t lymphocyte antigen-4 blockade enhances antitumor immunity by stimulating
  melanoma-specific t-cell motility, Cancer Immunology Research 2~(10) (2014)
  970--980.

\bibitem{linhares2019therapeutic}
A.~D.~S. Linhares, C.~Battin, S.~Jutz, J.~Leitner, C.~Hafner, J.~Tobias,
  U.~Wiedermann, M.~Kundi, G.~J. Zlabinger, K.~Grabmeier-Pfistershammer,
  et~al., Therapeutic pd-l1 antibodies are more effective than pd-1 antibodies
  in blocking pd-1/pd-l1 signaling, Scientific Reports 9~(1) (2019) 1--9.

\bibitem{chen2018sequential}
Y.-J. Chen, W.-C. Huang, S.-Y. Liu, C.-C. Ko, Sequential blockade of pd-1 and
  pd-l1 causes fulminant cardiotoxicity: From case report to mice model
  validation, Annals of Oncology 29 (2018) viii431.

\bibitem{shen2018efficacy}
X.~Shen, B.~Zhao, Efficacy of pd-1 or pd-l1 inhibitors and pd-l1 expression
  status in cancer: meta-analysis, Bmj 362 (2018).

\bibitem{basanta2012investigating}
D.~Basanta, J.~G. Scott, M.~N. Fishman, G.~Ayala, S.~W. Hayward, A.~R.
  Anderson, Investigating prostate cancer tumour--stroma interactions: clinical
  and biological insights from an evolutionary game, British Journal of Cancer
  106~(1) (2012) 174--181.

\bibitem{szeto2019integrative}
G.~L. Szeto, S.~D. Finley, Integrative approaches to cancer immunotherapy,
  Trends in Cancer 5~(7) (2019) 400--410.

\bibitem{altrock2015mathematics}
P.~M. Altrock, L.~L. Liu, F.~Michor, The mathematics of cancer: integrating
  quantitative models, Nature Reviews Cancer 15~(12) (2015) 730--745.

\bibitem{brady2019mathematical}
R.~Brady, H.~Enderling, Mathematical models of cancer: when to predict novel
  therapies, and when not to, Bulletin of Mathematical Biology 81~(10) (2019)
  3722--3731.

\bibitem{garcia2020cancer}
V.~Garcia, S.~Bonhoeffer, F.~Fu, Cancer-induced immunosuppression can enable
  effectiveness of immunotherapy through bistability generation: a mathematical
  and computational examination, Journal of Theoretical Biology 492 (2020)
  110185.

\bibitem{brown2020assessing}
M.~E. Brown, D.~Bedinger, A.~Lilov, P.~Rathanaswami, M.~V{\'a}squez, S.~Durand,
  I.~Wallace-Moyer, L.~Zhong, J.~H. Nett, I.~Burnina, et~al., Assessing the
  binding properties of the anti-pd-1 antibody landscape using label-free
  biosensors, PLoS ONE 15~(3) (2020) e0229206.

\bibitem{loewe1926kombinationswirkungen}
S.~t. Loewe, H.~Muischnek, {\"U}ber kombinationswirkungen, Naunyn-Schmiedebergs
  Archiv f{\"u}r experimentelle Pathologie und Pharmakologie 114~(5-6) (1926)
  313--326.

\bibitem{depillis2014modeling}
L.~DePillis, A.~Eladdadi, A.~Radunskaya, Modeling cancer-immune responses to
  therapy, Journal of Pharmacokinetics and Pharmacodynamics 41~(5) (2014)
  461--478.

\bibitem{de2005validated}
L.~G. de~Pillis, A.~E. Radunskaya, C.~L. Wiseman, A validated mathematical
  model of cell-mediated immune response to tumor growth, Cancer Research
  65~(17) (2005) 7950--7958.

\bibitem{wilson2012mathematical}
S.~Wilson, D.~Levy, A mathematical model of the enhancement of tumor vaccine
  efficacy by immunotherapy, Bulletin of Mathematical Biology 74~(7) (2012)
  1485--1500.

\bibitem{yamamoto2016tumour}
Y.~Yamamoto, C.~P. Offord, G.~Kimura, S.~Kuribayashi, H.~Takeda, S.~Tsuchiya,
  H.~Shimojo, H.~Kanno, I.~Bozic, M.~A. Nowak, et~al., Tumour and immune cell
  dynamics explain the psa bounce after prostate cancer brachytherapy, British
  Journal of Cancer 115~(2) (2016) 195--202.

\bibitem{serre2016mathematical}
R.~Serre, S.~Benzekry, L.~Padovani, C.~Meille, N.~Andr{\'e}, J.~Ciccolini,
  F.~Barlesi, X.~Muracciole, D.~Barbolosi, Mathematical modeling of cancer
  immunotherapy and its synergy with radiotherapy, Cancer Research 76~(17)
  (2016) 4931--4940.

\bibitem{castiglione2007cancer}
F.~Castiglione, B.~Piccoli, Cancer immunotherapy, mathematical modeling and
  optimal control, Journal of Theoretical Biology 247~(4) (2007) 723--732.

\bibitem{owens2020modelling}
K.~L. Owens, I.~Bozic, Modelling car t-cell therapy with patient
  preconditioning, bioRxiv (2020).

\bibitem{kimmel2020response}
G.~J. Kimmel, F.~L. Locke, P.~M. Altrock, Response to car t cell therapy can be
  explained by ecological cell dynamics and stochastic extinction events,
  bioRxiv (2020) 717074.

\bibitem{kirschner1998modeling}
D.~Kirschner, J.~C. Panetta, Modeling immunotherapy of the tumor--immune
  interaction, Journal of Mathematical Biology 37~(3) (1998) 235--252.

\bibitem{eftimie2011interactions}
R.~Eftimie, J.~L. Bramson, D.~J. Earn, Interactions between the immune system
  and cancer: a brief review of non-spatial mathematical models, Bulletin of
  Mathematical Biology 73~(1) (2011) 2--32.

\bibitem{lakatos2020evolutionary}
E.~Lakatos, M.~J. Williams, R.~O. Schenck, W.~C. Cross, J.~Househam, L.~Zapata,
  B.~Werner, C.~Gatenbee, M.~Robertson-Tessi, C.~P. Barnes, et~al.,
  Evolutionary dynamics of neoantigens in growing tumors, Nature Genetics
  52~(10) (2020) 1057--1066.

\bibitem{west2019immune}
J.~West, M.~Robertson-Tessi, K.~Luddy, D.~S. Park, D.~F. Williamson, C.~Harmon,
  H.~T. Khong, J.~Brown, A.~R. Anderson, The immune checkpoint kick start:
  Optimization of neoadjuvant combination therapy using game theory, JCO
  Clinical Cancer Informatics 3 (2019) 1--12.

\bibitem{sui2018anti}
H.~Sui, N.~Ma, Y.~Wang, H.~Li, X.~Liu, Y.~Su, J.~Yang, Anti-pd-1/pd-l1 therapy
  for non-small-cell lung cancer: toward personalized medicine and combination
  strategies, Journal of Immunology Research 2018 (2018).

\bibitem{sato2020rationale}
H.~Sato, N.~Okonogi, T.~Nakano, Rationale of combination of anti-pd-1/pd-l1
  antibody therapy and radiotherapy for cancer treatment, International Journal
  of Clinical Oncology 25~(5) (2020) 801--809.

\bibitem{page2019two}
D.~B. Page, H.~Bear, S.~Prabhakaran, M.~E. Gatti-Mays, A.~Thomas, E.~Cobain,
  H.~McArthur, J.~M. Balko, S.~R. Gameiro, R.~Nanda, et~al., Two may be better
  than one: Pd-1/pd-l1 blockade combination approaches in metastatic breast
  cancer, NPJ Breast Cancer 5~(1) (2019) 1--9.

\bibitem{wang2019silico}
H.~Wang, O.~Milberg, I.~H. Bartelink, P.~Vicini, B.~Wang, R.~Narwal, L.~Roskos,
  C.~A. Santa-Maria, A.~S. Popel, In silico simulation of a clinical trial with
  anti-ctla-4 and anti-pd-l1 immunotherapies in metastatic breast cancer using
  a systems pharmacology model, Royal Society Open Science 6~(5) (2019) 190366.

\bibitem{milberg2019qsp}
O.~Milberg, C.~Gong, M.~Jafarnejad, I.~H. Bartelink, B.~Wang, P.~Vicini,
  R.~Narwal, L.~Roskos, A.~S. Popel, A qsp model for predicting clinical
  responses to monotherapy, combination and sequential therapy following
  ctla-4, pd-1, and pd-l1 checkpoint blockade, Scientific Reports 9~(1) (2019)
  1--17.

\bibitem{gong2017computational}
C.~Gong, O.~Milberg, B.~Wang, P.~Vicini, R.~Narwal, L.~Roskos, A.~S. Popel, A
  computational multiscale agent-based model for simulating spatio-temporal
  tumour immune response to pd1 and pdl1 inhibition, Journal of the Royal
  Society Interface 14~(134) (2017) 20170320.

\end{thebibliography}

\end{document}